# ARTIFICIAL GAUGE FIELDS
## with ultracold atoms

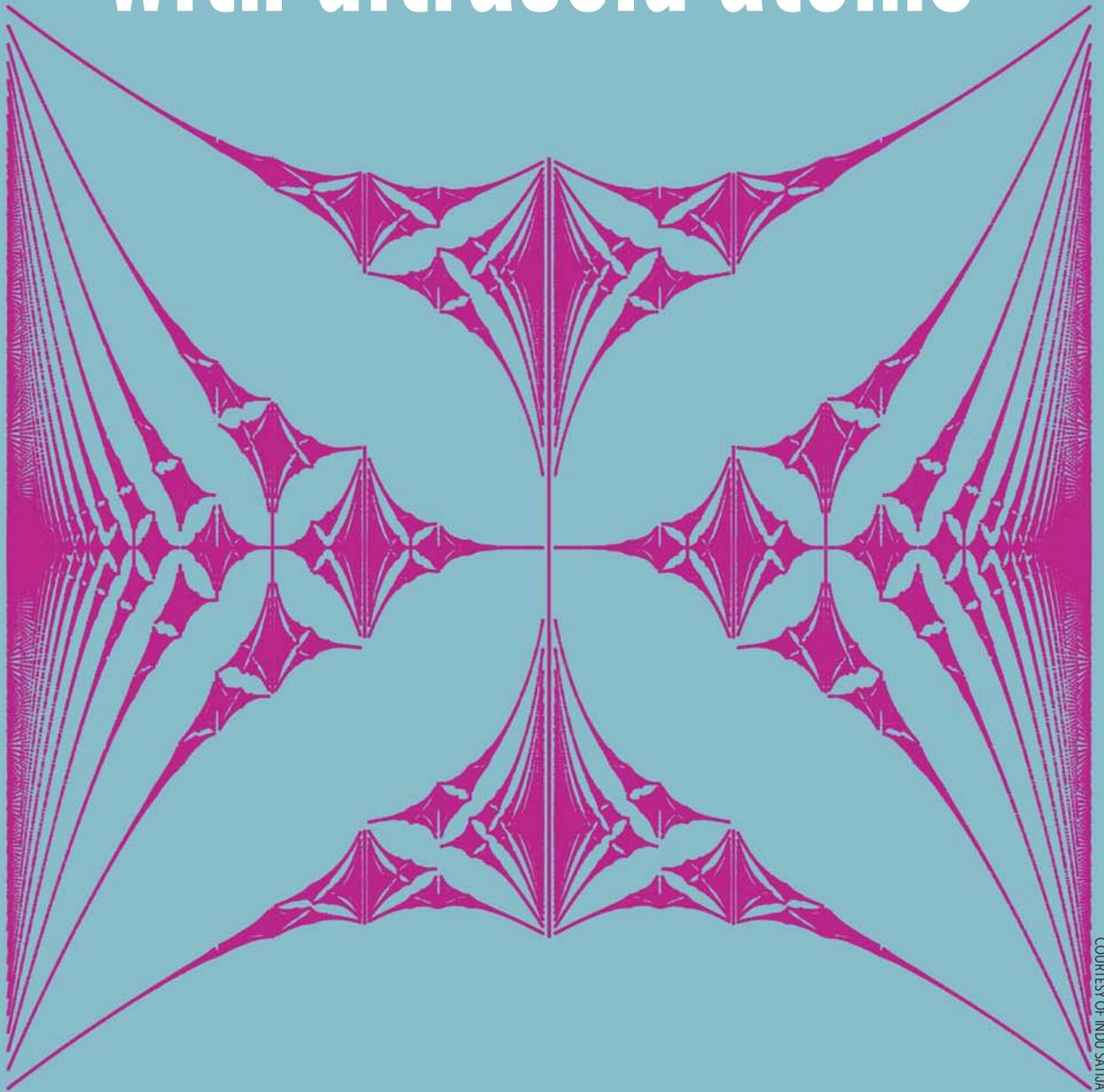

COURTESY OF INDU SATIJA

**Victor Galitski** and **Ian Spielman** are both fellows at the Joint Quantum Institute, in College Park, Maryland; the research partnership is between NIST, where Spielman is a NIST fellow, and the University of Maryland, where he and Galitski are professors. **Gediminas Juzeliūnas** is a research professor at the Institute of Theoretical Physics and Astronomy at Vilnius University in Lithuania.

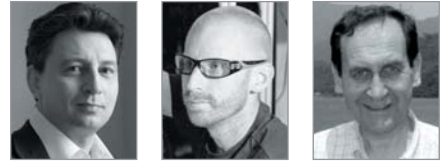

Victor Galitski, Gediminas Juzeliūnas, and Ian B. Spielman

**Suitable combinations of laser beams can make neutral atoms behave like electrons in a magnetic field.**

Gauge fields are ubiquitous in nature. In the context of quantum electrodynamics, you may be most familiar with the photon, which represents the gauge field mediating electromagnetic forces. But there are also gluons, which mediate strong forces, and the W and Z particles, which mediate the weak forces. According to the standard model, those few gauge bosons, in fact, mediate all elementary interactions.

In materials physics, applied gauge fields in the form of laboratory magnetic fields are essential for realizing exotic quantum phenomena, such as the quantum Hall effect (see the article by Joseph Avron, Daniel Osadchy, and Ruedi Seiler, PHYSICS TODAY, August 2003, page 38). In a two-dimensional lattice, for instance, electrons that move through a periodic potential and a strong magnetic field are expected to exhibit a recursive, fractal energy spectrum known as Hofstadter's butterfly, pictured here.

Furthermore, strongly correlated systems can host new gauge fields or particles known as anyons, which obey unusual quantum statistics—not those of everyday fermions or bosons. This article explains how to create artificial gauge fields—made on demand in a controlled laboratory setting—by using ultracold atomic gases. The engineered gauge fields can lead to phenomena that naturally occur in other systems and to new phenomena that occur nowhere else in nature.

To understand the concept of a gauge field, one must confront a difficult and perplexing aspect of physics, in which the equations describing physical reality involve variables that are not measurable. Such equations occur even in elementary classical electrodynamics, where the magnetic vector potential **A**, satisfying **B** = ∇×**A**, is used in place of the magnetic field **B** to describe the familiar Lorentz force that curves charged particles around magnetic field lines. A static vector potential is not measurable by itself—only its curl, the actual magnetic field, is. Hence **A** is defined only up to an arbitrary curl-free function—allowing the so-called gauge choice—that keeps the observable electromagnetic forces intact. That reality cannot depend on an arbitrary function is an example of gauge invariance, a fundamental physical principle that requires all physical observables to be independent of the choice of a gauge.

The concept of gauge field is merely auxiliary in classical physics. Maxwell's equations and Newton's force law can optionally be reframed in the Hamiltonian or Lagrangian formalisms that replace physical fields with gauge potentials. Although the gauge potentials provide a convenient mathematical framework for solving some problems, a complete understanding of classical physics does not require the reframing.

In contrast, the gauge-field formulation is the only consistent mathematical description of quantum particles interacting with electromagnetic fields. Indeed, the central object in quantum mechanics, the wavefunction $\psi(\mathbf{r})$—often represented as a position-dependent complex function having both real and imaginary components—is not directly observable. The wavefunction's overall complex phase factor $e^{i\varphi}$ can be changed with no physical consequence. For charged particles, that ambiguity in the definition of a wavefunction is tied one-to-one to the ambiguity in the choice of a gauge for the electromagnetic field.

Gauge ambiguity is not a mathematical curiosity. It's an intrinsic property of quantum mechanics and a source of many important and peculiar quantum effects, such as the Aharonov–Bohm effect (see the article by Herman Batelaan and Akira Tonomura, PHYSICS TODAY, September 2009, page 38) and topological phases of matter, the focus of the 2016 Nobel Prize in Physics (see PHYSICS TODAY, December 2016, page 14).

## Ultracold atoms

Although gauge fields and their associated forces abound in physics, their properties cannot be fully controlled. For example, the charge of an electron is a fundamental constant, and the Lorentz force is a law of nature; both are nonnegotiable. And yet gauge-field physics can be simulated in tabletop experiments by using ultracold atoms.

A cloud of ultracold atoms, held at temperatures typically between hundreds of picokelvins to tens of microkelvins, is up to a million times thinner than air. The atoms live in ultrahigh vacuum, isolated from their environment and trapped by optical or magnetic forces. Each aspect of their physical description must be assembled from quantum mechanical building blocks.

The energy of the atoms' interactions with each other is feeble, typically on the order of $10^{-31}$ joules. Yet the gases can still form strongly interacting quantum systems. For example, neutral atoms trapped in an optical lattice—the standing waves formed by mutually interfering laser beams—tunnel from one site to another, just as electrons do in a crystalline solid, and repel each other whenever they share the same lattice site.

The energy balance between the atoms' tunneling and their interactions can be adjusted by changing the intensity of the interfering lasers. The change can drive the transition between an itinerant phase, in which atoms hop freely between lattice sites, and a crystalline phase, in which they are immobile, pinned to their sites. The change in intensity also provides a way to realize the Hubbard model in strongly correlated electronic materials (see the article by Gabriel Kotliar and Dieter Vollhardt, PHYSICS TODAY, March 2004, page 53). The Hubbard model and other standard theoretical models rarely occur in their pristine form in materials. To give such iconic and idealized models life in the laboratory, one can use cold atoms to forge a link between quantitative, precise atomic-physics experiments and many-body theory.

The response of a material to electromagnetic fields is perhaps the most versatile and informative way to probe its electronic phases. As simulators, though, cold atoms lack a crucial ingredient that makes such probing possible: electric charge. Because of their charge neutrality, individual atoms do not experience Lorentz forces in a magnetic field. It would therefore seem that a wide range of phenomena would be forever out of reach of cold-atom experiments. Fortunately, that's not so.

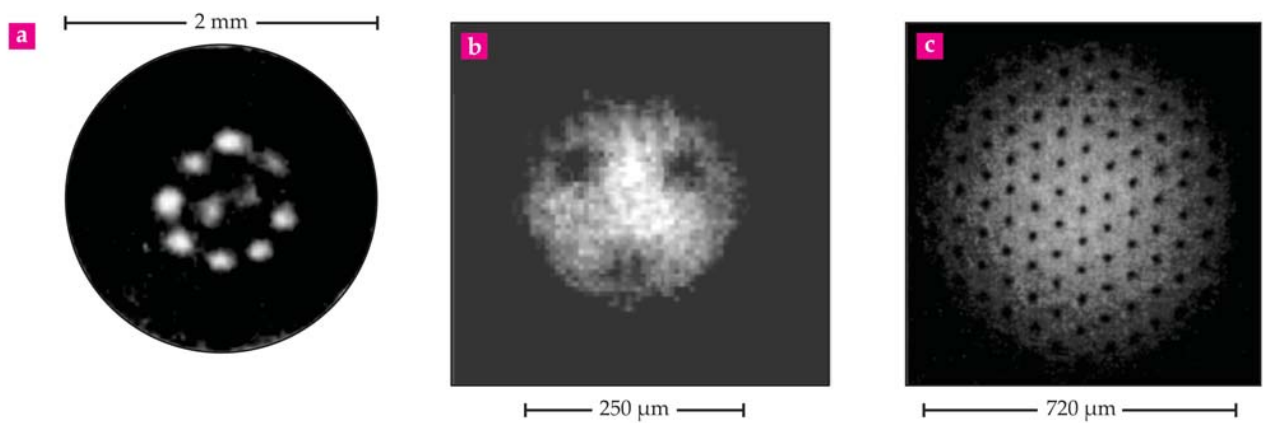

**FIGURE 1. THE APPEARANCE OF QUANTIZED VORTICES** in a rotating superfluid is a smoking gun for the presence of an artificial magnetic field. The density of vortices is in direct proportion to the artificial field's magnitude. **(a)** Quantized vortex arrays in superfluids were first imaged in liquid helium in 1979 at the University of California, Berkeley.[1] **(b)** They were later seen, in 2000, in a slowly rotating Bose–Einstein condensate (BEC) of rubidium.[2] **(c)** This 2003 image of a quickly rotating BEC of Rb captures a large array of vortices created at JILA.[4]

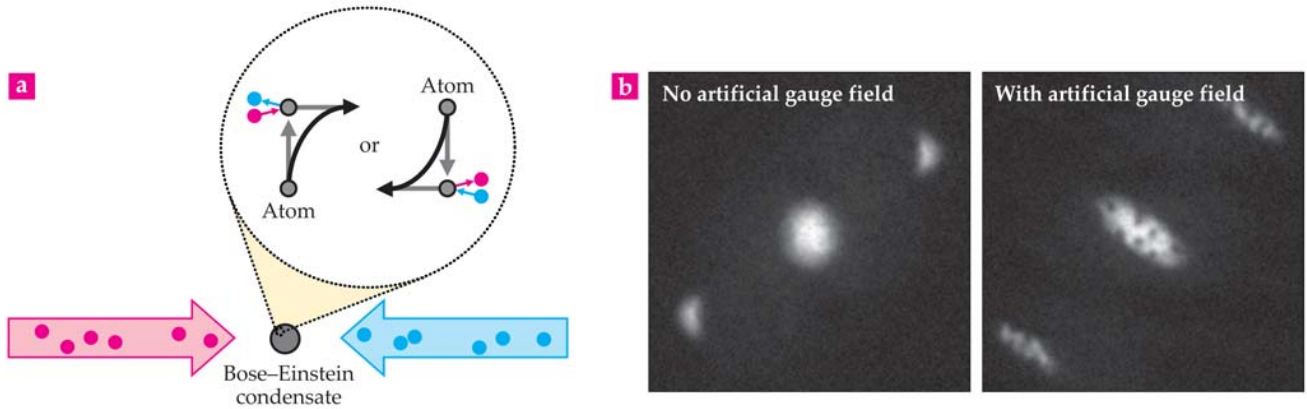

**FIGURE 2. STIMULATED RAMAN TRANSITIONS.** Two counterpropagating laser beams (pink and blue) incident on a Bose–Einstein condensate (BEC) can create artificial magnetic fields. **(a)** The beams drive stimulated Raman transitions in the atoms; each transition absorbs a photon from one beam and emits one into the other beam while imparting a momentum kick to the atom. As an atom is moving upward, say, it also moves right from the recoil. The reverse happens for atoms moving downward. The kicks simulate a transverse, velocity-dependent Lorentz force. **(b)** An atomic BEC with an artificial gauge field turned off (left) and turned on (right). The appearance of quantized vortices mark the presence of the artificial gauge field. (Adapted from ref. 5.)

## Rotation as simulation

The first artificial magnetic field experiments using cold atoms exploited the equivalence between the Lorentz force in a uniform magnetic field and the Coriolis force in a spatially rotating frame. The equivalence may be most familiar in the context of the Foucault pendulum, whose axis of oscillation slowly rotates at an angular velocity $\omega$. The reason for the rotation is simple: A particle traveling linearly with velocity $v$ in a stationary frame undergoes curved motion in a rotating frame from the Coriolis force $\mathbf{F}_C \propto \mathbf{v} \times \boldsymbol{\omega}$, just as a charged particle follows a circular cyclotron orbit in a uniform magnetic field from the Lorentz force $\mathbf{F}_L \propto \mathbf{v} \times \mathbf{B}$.

One of the most beautiful and direct manifestations of superconductivity is the formation of vortices—sharply localized quanta of circulation of magnetic flux or angular momentum. Quantized vortices were predicted for superfluid helium by Lars Onsager in 1947 and by Richard Feynman in 1955. Russian theorist Alexei Abrikosov extended the prediction to superconductors in 1957 and showed that interactions between the vortices ordered them into a regular array, the so-called Abrikosov lattice. The achievement earned him a Nobel Prize in Physics almost 50 years later (see PHYSICS TODAY, December 2003, page 21).

Vortices can be induced in a superfluid by the effective magnetic field present in rotating systems. In 1979 Richard Packard and colleagues first imaged a vortex array, as shown in figure 1, in helium by using a rotating cryostat.[1] Two decades later, Jean Dalibard and colleagues found much the same thing—a small cluster of three uniformly spaced vortices—in atomic Bose–Einstein condensates (BECs) of rubidium-87 stirred by a focused laser beam.[2] The observation provided strong evidence for superfluidity in BECs.

The groups of Wolfgang Ketterle at MIT and Eric Cornell at JILA made technical improvements that led to observations of large vortex arrays—a qualitative leap beyond the earlier results found in liquid He and the atomic BECs.[3,4] Those large Abrikosov lattices set the stage for studying the dynamics of vortex crystals during melting and in other nonequilibrium settings.

As exciting as the experiments on rotating superfluids have been, though, they only probed phenomena observable in weak gauge fields. An important figure of merit in many-body

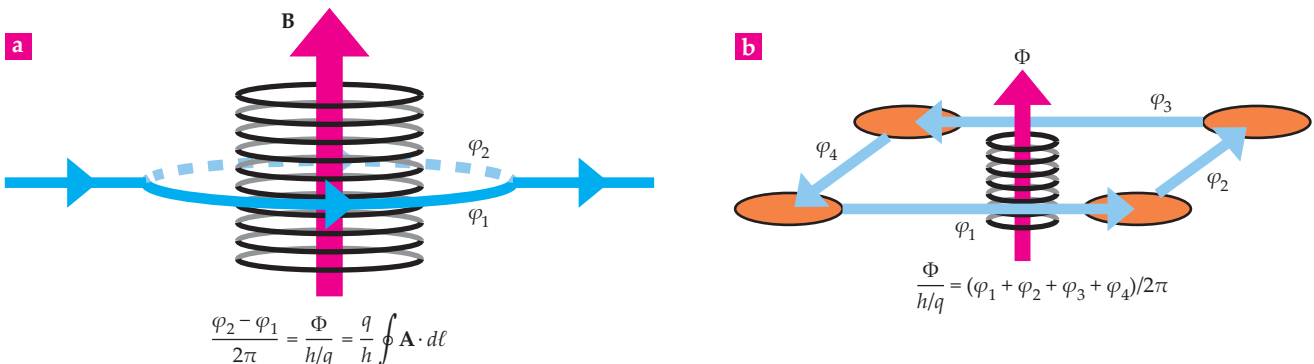

$$\frac{\varphi_2 - \varphi_1}{2\pi} = \frac{\Phi}{h/q} = \frac{q}{h}\oint \mathbf{A} \cdot d\ell$$

$$\frac{\Phi}{h/q} = (\varphi_1 + \varphi_2 + \varphi_3 + \varphi_4)/2\pi$$

**FIGURE 3. PHASE ENGINEERING.** Quantum mechanical particles experience magnetic fields by acquiring Aharonov–Bohm (AB) phases. **(a)** In an AB interferometer, a magnetic field **B** is confined to the interior of a solenoid and is zero outside it. Nevertheless, a particle that travels on opposite sides of the solenoid will acquire a phase difference proportional to the enclosed magnetic flux $\Phi$. That flux equals the line integral of the vector potential **A** around the solenoid. The charge is specified by $q$. **(b)** Particles that tunnel in a lattice experience an applied magnetic field in much the same way as in the AB effect. Their tunneling motion from one site to another acquires a so-called Peierls phase, such that the sum of the phases around a closed loop equals the enclosed flux.

systems is the ratio of the radius of the minimum cyclotron orbit to the average interparticle separation. The smaller the ratio, the stronger the magnetic field effects and the closer one gets to probing the difficult and exciting physics of strongly correlated topological phases of matter. Those phases include the fractional quantum Hall states, collective states of matter that occur in ultrastrong fields and in which electrons behave as if their elementary charge is a fraction of their actual charge. (See the article by Jainendra Jain, PHYSICS TODAY, April 2000, page 39.) Unfortunately, the range of rotation-induced artificial magnetic fields are far from that interesting territory. New approaches were required to reach larger artificial fields.

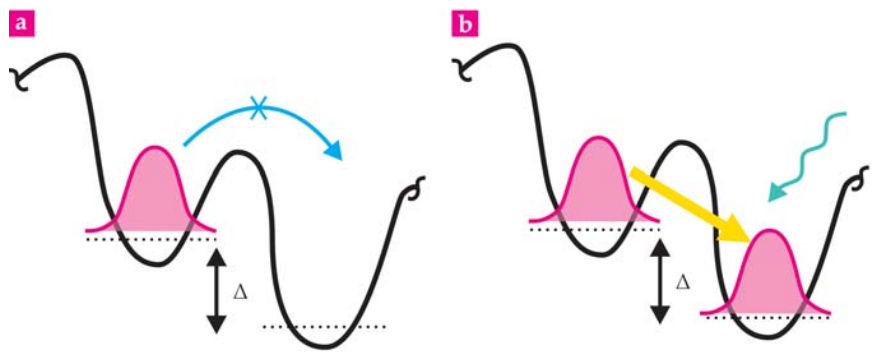

**FIGURE 4. LASER-ASSISTED TUNNELING.** The native tunneling in an optical lattice does not produce the phase differences required to emulate a gauge field. But the phases can be created in a two-step process. **(a)** When a potential gradient of energy Δ is applied to a lattice, the energy states in adjacent sites go out of resonance with each other, which blocks a localized atomic wavepacket (pink) from tunneling. **(b)** Additional laser fields illuminating the lattice can provide the required energy to reestablish tunneling (yellow arrow). The applied fields also imprint a local optical phase on the atom's wavefunction as it moves between lattice sites.

### Laser-induced gauge fields

The first new approach in the laboratory used an atomic BEC illuminated by a pair of counterpropagating laser beams, each having a distinct wavelength $\lambda$ (figure 2a). Most atoms, including those in a BEC, have several spin states that are distinguished by a combination of orbital, electron, and nuclear spin degrees of freedom. When the laser beams strike the BEC, they become quantum mechanically coupled to the spin states of individual atoms. During each interaction, an atom's spin flips, a process accompanied by the exchange of a photon from one beam to the other. The exchange imparts to the atom a momentum kick of $2h/\lambda$, which changes the atomic velocity by about 1 cm/s. The BEC atoms are so cold that even that tiny change exceeds the thermal velocity of the system.

The recoil of the atoms depends on their vertical motion and slight differences in the beam wavelengths. As an atom moves upward, for instance, it is stimulated to emit a photon into the leftward-moving beam (blue) in figure 2 and, as a result, acquires a kick to the right. The change in momentum emulates the transverse response of a Lorentz force, which moves the atom in circles, to use the classical analogy. An artificial magnetic field is created.

Although that intuitive picture alludes to forces, the quantum system is mathematically better described by the appearance of an electromagnetic vector potential—a gauge field.[5–7] In 2009, a team including one of us (Spielman) demonstrated the technique at NIST.[5] The artificial magnetic field was marked by the presence of vortices, as shown in figure 2b, but in the laboratory frame rather than in a rotating frame.

In the final analysis, the experimental scheme in figure 2 turned out to be well suited for creating elongated trapped-atom geometries but not for creating extended systems with large fields. Even so, the pioneering work opened the door for experiments that now do operate at large fields.

### The Aharonov–Bohm effect

In quantum mechanical systems, gauge fields are inseparably intertwined with the wavefunction's phase. The connection is dramatically evident in the Aharonov–Bohm (AB) effect, shown in figure 3a, for charged quantum particles moving about an infinite solenoid. The magnetic field is zero outside the rings of the solenoid, where the particles actually move. But even though they never experience a magnetic field, the particles still respond to the electromagnetic vector potential, which, unlike the magnetic field, necessarily extends outside the solenoid.

Upon completing a closed loop, each particle's wavefunction acquires an additional phase—the AB phase—which is proportional to the magnetic flux enclosed by the solenoid. By Stokes's theorem, that flux is equal to the line integral of the vector potential around the loop traversed by the particle. Hence the acquired phase is a direct consequence of the vector potential, not the magnetic field itself.

One can see that connection in a lattice in which atoms tunnel between adjacent sites. The AB phase acquired by an atom as it encircles a square plaquette in the lattice, as shown in figure 3b, is just the sum of the four phases gained on the associated links.

### Gauge fields in optical lattices

Extending that one plaquette to a larger array would produce a 2D square lattice with a constant magnetic flux in each unit cell. But how can the tunneling phases be created in the laboratory without the real magnetic field of a solenoid? The first theoretical proposal was to make an artificial gauge field that uses "laser-induced tunneling" of atoms between sites in an optical lattice.[8] In the proposal, the atoms don't tunnel through the barrier on their own volition; rather, they are pushed through it with additional laser fields.

The concept is illustrated in figure 4, using just two lattice sites: Imagine a conventional optical lattice that is tilted—with one site higher in energy than the other. The tilting takes the atoms in the two sites out of resonance with each other and prevents tunneling. Two light fields whose photon energies differ by the tilt energy then provide just the right amount of energy to link the states together.

In addition to reestablishing tunneling between sites, the laser-induced tunneling also imprints a position-dependent optical phase onto the wavefunction of the atoms. The phases emulate the quantum mechanical phases picked up by electrons tunneling between the lattice sites of a crystal in a real magnetic field.[7] The optical phases that accumulate around each square plaquette in the lattice are equal and emerge from a uniform artificial magnetic field.

Unlike in naturally occurring solids, where the magnetic flux through a plaquette is much smaller than the flux quantum $h/q$, where $q$ is the magnitude of the relevant charge, very large laser-induced fluxes can be produced in an optical lattice. Figure 5a shows one experimental demonstration,[9] in which an atomic wavepacket prepared in a single site of an optical lattice undergoes cyclotron-like motion around a square plaquette; its center-of-mass motion, from dark green to light green, is plotted over 2 ms. A similar scheme was experimentally implemented by Ketterle's group around the same time.[10] Nontrivial phases have also been imprinted by literally shaking the optical lattices,[11] but in a way that does not correspond to a uniform magnetic field.

### Synthetic dimensions

An essential step in creating gauge fields with laser-assisted tunneling is to suppress the native tunneling of atoms between lattice sites. In the preceding section, we took the natural approach of using a tilted lattice to do the job. But that's not the only solution. Five years ago, a group of researchers, including two of us (Spielman and Juzeliūnas), proposed a scheme using transitions between different spin states of the BEC atomic ground state to create "synthetic dimensions."[12]

Lattice sites, according to that approach, need not be in different places in space. Dimension can also refer to the connectivity, or number of independent states, into which the atoms can tunnel. In addition to a real dimension, along which motion corresponds to a displacement in space, there's a synthetic dimension that corresponds to the internal spin states of ultracold atoms. When lasers couple the spin states together, the resulting "motion" between the spin states is accompanied by an optical phase that gives rise to a uniform flux.

> In addition to reestablishing tunneling between sites, the laser-induced tunneling also imprints a position-dependent optical phase onto the wavefunction of the atoms.

The concept, which Spielman and others subsequently implemented in 2015, offers new opportunities for control and measurement.[13] For one thing, it's easy to detect individual synthetic lattice sites using standard spin-selective detection techniques. Also in 2015 Leonardo Fallani's group concurrently demonstrated the utility of synthetic dimensions in a degenerate Fermi gas of ytterbium.[14]

Figure 5b presents the experimental trajectories followed by BEC atoms on a long, thin virtual strip, whose real dimension is space, plotted horizontally, and whose synthetic dimension is the atoms' magnetic quantum numbers $m = 1$, 0, and −1, plotted vertically.[13] By preparing the atoms on the strip's edges and then allowing them to move in the artificial magnetic field, the researchers observed them following "skipping orbits" along the top and bottom edges. Each time an atom struck the hard-wall potential along an edge, it reflected and began following yet another cyclotron orbit.

### To be non-abelian

As we've seen, artificial gauge fields make charge-neutral particles curve as if they were under the influence of a Lorentz force. But in quantum systems, the rabbit hole is deeper: As we have noted, quantum particles have internal degrees of freedom, like the spin-up and spin-down of an electron.

That simple addition opens the door for new kinds of so-called non-abelian gauge fields, in which the effective Lorentz force can depend on the magnitude and sign of the spin or even drive changes to the spin state as the particle moves. (Such a non-abelian gauge field is distinct from the non-abelian quasiparticles that are potentially relevant for quantum computation. See the article by Nick Read, PHYSICS TODAY, July 2012, page 38.) In essence, the Aharonov–Bohm phases $\varphi$ of figure 3a would each be replaced by a matrix that can give different phases for the different spin states or that can even change the spin state entirely.

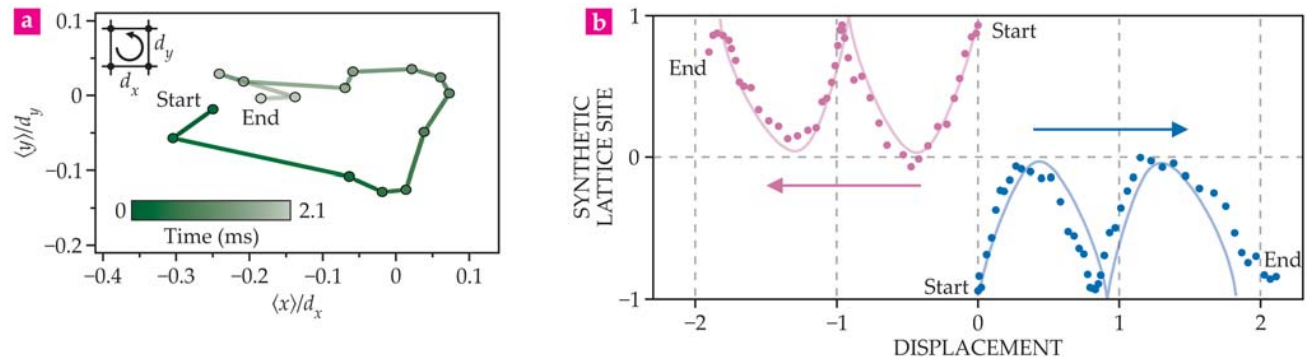

**FIGURE 5. DYNAMICS IN AN ARTIFICIAL GAUGE FIELD.** Atoms that move from one optical lattice site to another under laser-assisted tunneling circle through the sites, as in figure 3b. **(a)** One experiment measured the center-of-mass positions of atoms (dots) $\langle x \rangle$ and $\langle y \rangle$, scaled to the lattice constants $d_x$ and $d_y$, as the atoms move among the four sites (inset) over 2.1 ms. (Adapted from ref. 9.) **(b)** In another experiment, the positions (dots) of BEC atoms are mapped along the edges of a long, thin virtual strip. The horizontal dimension, scaled to the lattice constants, is real, and the narrow dimension is synthetic, formed by the atoms' spin states 1, 0, and −1. The edge currents to the left (pink) and right (blue) evolve in time with opposite velocities and are the result of the "skipping orbits," in which the atoms mimic cyclotron motion and periodically reflect from the strip's edges. (Adapted from ref. 13.)

Spin–orbit coupling in a material is an interaction between the electron's momentum and its spin. In many cases, the interaction is equivalent to a non-abelian gauge field. Spin–orbit coupling is responsible for various interesting physical phenomena—from the spin Hall effect to Majorana fermions and topological insulators (see the box on this page and PHYSICS TODAY, March 2011, page 20). By extending the ideas discussed above, certain types of non-abelian gauge fields have also seen the light of day in cold-atom experiments.[7]

## Applications

The experiments described in this article serve as quantum simulations to better understand the toy models that approximately describe real materials. Vortex physics is an ideal target for quantum simulation. Because the motion of vortices is a leading source of dissipation in superconductors, understanding them has wide-ranging real-world impact, from high-field superconducting magnets used in medical magnetic resonance imaging to magnetic levitating trains.

Much is still unknown about quantum vortices: How do large collections of them interact and evolve? (See, for example, PHYSICS TODAY, January 2017, page 19.) When do their positions become pinned to the disorder potential? How do we understand the flow of angular momentum in a material with vortices, in analogy to the flow of electrons in a metal? Cold-atom superfluids in the presence of an artificial magnetic field serve as a medium for exploring those questions.

Artificial gauge fields provide new techniques for realizing topological states of matter (see the Quick Study by Mohammad Hafezi and Jake Taylor, PHYSICS TODAY, May 2014, page 68). Topological insulators and the integer and fractional quantum Hall effects were discovered and understood in conventional material systems, but the practical limitations of material systems hinder the ability to create new types of topological matter. For example, although many 2D models host anyon excitations, their only known physical manifestation is the fractional quantum Hall effect. Three-dimensional counterparts—interacting topological insulators—are beyond the reach of current experiment and theory. Cold atoms with artificial magnetic fields provide a realistic system experimentalists can use to engineer and observe their exotic states.

Artificial gauge fields can also host completely new physics that have no analogues elsewhere in nature. One example is spin-½ bosons made with spin–orbit coupling. Recall that bosons normally have integer spin and fermions have half-integer spin. As in quantum Hall systems, spin-½ bosons would boast a massively degenerate ground state and be ideally suited for creating strongly correlated topological matter.[16]

The gauge fields present in high-energy physics are, like the photon, dynamical gauge fields. Aspects of those dynamical gauge fields may be modeled with time-dependent artificial gauge fields. A dynamical gauge field samples all possible configurations of the associated classical field. In certain cases, quantum fluctuations in a dynamical gauge field have average properties similar to that of noise added to a classical gauge field. Sampling those fluctuations can be experimentally modeled using atoms in optical lattices coupled to a synthetic gauge field to which laboratory-controlled noise has been added. That would be a good first step to simulating lattice gauge theories in the low-temperature limit.

> ## TOPOLOGICAL MATTER
>
> Band insulators are crystalline materials with an energy gap between the materials' occupied valence bands and unoccupied conduction bands. A single excitation produces a mobile electron free to move in the crystal lattice. The charge and quantum statistics of that excitation are the same as those of the material's constituent electrons.
>
> Topological insulators (TIs) are a broad class of band insulators, distinguished by different windings of the phase of the atom's eigenstates for different crystal momenta in the lattice. Although excitations deep inside TIs separate, much like conventional insulators, into valence and conduction bands, TIs form new conducting states at crystal boundaries. (See the article by Xiao-Liang Qi and Shou-Cheng Zhang, PHYSICS TODAY, January 2010, page 33.)
>
> Strongly correlated materials give new excitations called anyons. As Frank Wilczek put it, "The statistics of these objects, like their spin, interpolates continuously between the usual boson and fermion cases."[15] Often referred to as fractionalized quasiparticles, anyons occur in strongly correlated topologically ordered states.
>
> Such strongly correlated states are theoretically present in various quantum spin models. But laboratory realizations of TIs are few: They have been seen experimentally only in fractional quantum Hall systems, which require highly restrictive experimental conditions, such as ultralow temperatures, high magnetic fields, and ultraclean samples. Strongly interacting ultracold atoms subject to artificial gauge fields are one promising avenue for realizing those fractional states of matter.

*For financial support, we thank NIST, the US Army Research Office, the US Air Force Office of Scientific Research, NSF through its Physics Frontier Center at the Joint Quantum Institute at the University of Maryland, and the Lithuanian Research Council. Victor Galitski and Ian Spielman also appreciate the support of Paul Baker and Peter Reynolds.*

PT